\documentstyle[preprint,epsfig,aps]{revtex}

\begin{document}
\draft
\title{Attraction between like-charged colloidal particles induced by a surface
:\\ a density - functional analysis}
\author{David Goulding and Jean-Pierre Hansen}
\address{Department of Chemistry, University of Cambridge \\
Lensfield Road, Cambridge CB2 1EW (UK)}
\date{\today}
\maketitle
\begin{abstract}
We show that the first non-linear correction to the linearised Poisson-Boltzman
n (or DLVO) theory of effective pair interactions between charge-stabilised, co
lloidal particles near a charged wall leads to an attractive component of entro
pic origin. The position and depth of the potential compare favourably with rec
ent experimental measurements
\end{abstract}

Recent direct measurements confirm the overall validity of the classic Derjagui
n-Landau-Verwey-Overbeek (DLVO) effective pair potential between charge-stabili
zed colloidal particles in the bulk of an aqueous dispersion \cite{1}, but give
 convincing evidence for an attractive component to the interaction when two co
lloidal particles are confined in a slit \cite{2,3} or close to a planar wall \
cite{4}. In this letter we use a density functional analysis \cite{5} to show t
hat such an attraction can be due to non-linear corrections to linearised Poiss
on-Boltzmann (PB) theory which forms the basis of the DLVO potential.

Mesoscopic colloidal particles carrying ionizable radicals on their surface build up electric double -layers of microscopic co and counterions when dispersed in water. The effective forces between these ``dressed" particles or polyions derive from a sum of direct and induced interactions. The direct interactions include excluded volume and Coulomb repulsions between the bare surface charges, and van der Waals attractions; the latter are generally quite negligible,  except at very high salt concentrations corresponding to the strong screening regime \cite{6}. The indirect interactions are induced by the locally inhomogeneous distribution of microions, which screen the bare Coulomb repulsion; these induced interactions are identified with the free energy of the microions in the "external" field due to any given configuration of polyions, and hence have electrostatic and entropic contributions. In practice the free energy of the microions, which depends parametrically on the positions $\{\bf R \rm_i\}$ of the N  polyions (assumed here to be spherical) is calculated within density functional theory (DFT) by minimising an appropiate free energy functional F with respect to the local densities $\rho_\alpha(\bf r \rm)$ of the co and counterions. To be specific, consider polyions of radius R, carrying a total positive charge $Ze$ ($Z>0$), and monovalent coions (local density, $\rho_+(\bf r \rm))$ and counterions ($\rho_-(\bf r \rm)$). Z must be regarded as an effective charge much smaller than the bare structual charge. The ``renormalized'' charge accounts for the shortcomings of linearised PB theory (as used later in this paper) at very short distances from the polyion surfaces \cite{ACGMPH}. Correlations between microions will be neglected, which amounts to working within the framework of mean-field, or Poisson-Boltzmann theory. The solvent(water) is merely treated as a continium of dielectric constant $\epsilon$ (``primitive" model). Under these conditions the free energy functional reduces to the sum of ideal and electrostatic contributions:
\begin{eqnarray}
F[\rho_+,\rho_-]&=&F_{id}[\rho_+,\rho_-] + F_{el}[\rho_+,\rho_-] \\ \label{entropy}
F_{id}[\rho_+,\rho_-]&=&\sum_{\alpha=+,-} k_BT \int_\wp \rho_\alpha(\bf r \rm)\left[\log(\Lambda_\alpha^3\rho_\alpha(\bf r \rm))-1\right] d\bf r \\
F_{el}&=&\frac{1}{2}\int_\wp \rho_c(\bf r \rm) \Psi(\bf r \rm) d \bf r
\end{eqnarray}
where the $\Lambda_\alpha$ are irrelevant length scales, $\rho_c(\bf r\rm)$ is the total charge density:
\begin{equation}
\rho_c(\bf r\rm)=\sum_{i=1}^N \rho_{ext}^{(i)}(\bf r\rm) +e[\rho_+(\bf r\rm)-\rho_-(\bf r\rm)]
\end{equation}
$\rho_{ext}^{(i)}(\bf r\rm)$ is the ``external" charge density associated with polyion $i$, $\wp$ is the domain to which the dispersion is confined, and $\Psi(\bf r \rm)$ is the local electrostatic potential, which is related to $\rho_c(\bf r\rm)$ by Poisson's equation; the latter must be solved subject to appropiate boundary conditions on the surface bounding $\wp$.

If the $\rho_\alpha(\bf r\rm)$ are slowly varying, the ideal contribution $F_{id}$ may be expanded in a functional Taylor expansion around the mean (macroscopic) densities $n_\alpha$. If the expansion is truncated after second order, which amounts to linearisation of PB theory, the optimum density profiles (which minimise the free energy functional), and the resulting electrostatic potential, reduce to sums of terms (``orbitals") centred on each of the N polyions \cite{5}: 
\begin{equation}
\label{orbitals}
\Delta \rho_\alpha(\bf r\rm)=\rho_\alpha(\bf r\rm)-n_\alpha=\sum_{i=1}^N \rho_\alpha^{(i)}(\bf r \rm - \bf R \rm _i)
\end{equation}
where $\rho_\alpha^{(i)}(\bf r\rm)$ is constrained to vansh for $\mid \bf r \rm \mid < R$ (excluded volume). Apart from a structure-independant ``volume" term \cite{7} the total potential energy of the ``dressed" polyions reduces to a sum over all pairs of an effective pair potential:
\begin{equation}
\label{veff}
v_{eff}(\bf R \rm _i,\bf R \rm _j)=e\int_\wp \rho_{ext}^{(i)}(\bf r \rm)\Psi^{(j)}(\bf r \rm) d \bf r
\end{equation}
where $\Psi^{(j)}(\bf r \rm)$ is the electrostatic potential due to polyion j and its associated ``orbital" $\rho_+^{(j)}(\bf r \rm)-\rho_-^{(j)}(\bf r \rm)$. In the bulk, i.e.far from any surface bounding $\wp$, $v_{eff}$ reduces to the familiar DLVO potential \cite{8};
\begin{equation}
\label{DLVO}
v_{eff}(\bf R \rm _{ij})=Z_{eff}^2 e^2 \frac{exp\{-\kappa R_{ij}\}}{\epsilon R_{ij}}
\end{equation}
where $\kappa = 1/\lambda_D$ is the inverse Debye screening length, $\kappa =[4\pi(n_++n_-)e^2/\epsilon k_B T]^{1/2}$, and $Z_{eff}>Z$ is an effective polyion charge which accounts for polyion-microion excluded volume, resulting in $\rho_\alpha^{(j)}(\bf r \rm)=0$, $\mid \bf r \rm \mid < R$.

We have recently shown how $v_{eff}(\bf R \rm _{i},\bf R \rm _{j})$ is modified when the two polyions $i$ and $j$ are close to a planar surface, which separates the dispersion from a medium (e.g. glass) of dielectric constant $\epsilon'$ \cite{9} (see fig.1). Using an integral representaion of $\Psi^{(j)}(\bf r \rm)$ due to Stillinger \cite{10}, which takes proper account of image charges, we were able to show that within linearised PB theory, the effective repulsion between two polyions situated at an altitude h above the surface is enhanced relative to the bulk result (\ref{DLVO}), and decays like $1/R_{ij}^3$ at large (horizontal) separations $R_{ij}$. In fact, a careful asymptotic analysis of Stillinger's expression for $\Psi^{(j)}(s,z)$ (where $s$ is the radial coordinate parallel to the plane, and $z$ the vertical coordinate) shows that:
\begin{eqnarray}
\Psi^{(j)}(s,z)=\frac{Z_{eff} e}{\epsilon}\bigg[\frac{e^{-\kappa r}}{r}+\frac{e^{-\kappa r'}}{r'} &+&\frac{2\epsilon_r e^{-\kappa (2h+z)}}{s^3\kappa^2} + O(s^{-4})\bigg]
\end{eqnarray}
where all distances ($s$,$z$,$r=(s^2+z^2)^{1/2}$ and $r'=(s^2+(z+2h)^2)^{1/2}$) are relative to the centre of the polyion $j$, (or its image $j'$) and $\epsilon_r=\epsilon'/\epsilon$. The resulting $v_{eff}$ easily follows from eq.(\ref{veff}) and has the same asymptotic functional form.

It is thus clear that linearised PB theory cannot account for the attractive component of the effective pair potential between polyions, which is clearly apparent in recent experimental determinations of  $v_{eff}(\bf R \rm_i,\bf R \rm_j)$ in confined geometries [2-4] (for a possible interpretation of the experimental data, see however \cite{11}). The conclusion from linearised PB theory is independant of any surface charge carried by the planar boundary. To make progress, the full (non-linear) PB theory should be used. This can only be done numerically and represents a formidable task, in view of the abscence of simplifying symmetries. We hence resort to a perturbative approach, whereby the expansion of the ideal entropic terms in eq.(\ref{entropy}) around the uniform densities $n_\alpha$ is taken to third order, yielding:
\begin{eqnarray}
\label{idealexp}
F_{id}[\rho_+,\rho_-]&=&F_{id}(n_+,n_-)+k_BT\sum_\alpha \int_\wp \Delta \rho_\alpha(\bf r\rm)\Big[\log(\Lambda_\alpha^3n_\alpha)+ \frac{\Delta \rho_\alpha(\bf r\rm)}{2n_\alpha}\Big]d \bf r \rm \\ \nonumber &-&k_BT \sum_\alpha \int_\wp  \frac{\Delta \rho_\alpha^3(\bf r\rm)}{6 n_\alpha^2}d \bf r \rm+ O(\Delta\rho_\alpha^4)
\end{eqnarray}
linearised PB theory amounts to retaining only terms of up to quadratic order in $\Delta \rho_\alpha$, and leads back to the linear superposition (\ref{orbitals}) of ``orbitals" of the form given in \cite{9} and \cite{10}. In a density functional perturbation theory similar in spirit to that used by in references \cite{5} and  \cite{12} to determine triplet interactions in the bulk, this linear superposition of orbitals is then substituted into the second integral, involving cubic terms in $\Delta \rho_\alpha (\bf r \rm)$, in the functional expansion (\ref{idealexp}).

Since we are interested in the effect of a charged planar surface (carrying a uniform surface charge $\sigma$) on the effective pair potential between two polyions (say $i=1$ and $2$), the relevant superposition of ``orbitals", to be substituted into the perturbation term is;
\begin{equation}
\Delta \rho_\alpha (\bf r \rm)=\rho_\alpha^{(1)}+ \rho_\alpha^{(2)}+ \rho_\alpha^{(\sigma)}\ \ \ \ \ \ \ ;\ \alpha=+,-
\end{equation}
where $\rho_\alpha^{(\sigma)}(\bf r \rm)$ is the planar ``orbital", associated with the charged surface, as calculated within linearised PB theory, i.e.
\begin{equation}
\label{walldensity}
\rho_\alpha^{(\sigma)} (z)= \sigma \kappa e^{-\kappa (z+h)}
\end{equation} 
Introducing the local charge density in $\wp$:
\begin{eqnarray}\
\Delta \rho(\bf r \rm)&=& \Delta \rho_+(\bf r \rm)-\Delta \rho_-(\bf r \rm) \\ \nonumber
&=&\sum_i\left[\rho_+^{(i)}(\bf r \rm)-\rho_-^{(i)}(\bf r \rm)\right]\equiv \sum_i \rho^{(i)}(\bf r \rm)
\end{eqnarray}
the lowest-order non-linear correction to the effective pair-potential between polyions $1$ and $2$ reduces to:
\begin{eqnarray}
\label{expansion}
\Delta v_{eff}(\bf R \rm _1,\bf R \rm _2) &=& \frac{k_BT (n_- -n_+)}{(n_- + n_+)^3}\int_\wp \rho^{(1)}(\bf r \rm) \rho^{(2)}(\bf r \rm) \rho^{(\sigma)}(\bf r \rm) d \bf r \rm \\ \nonumber
&=& \frac{k_BT (n_--n_+)}{(n_-+n_+)^3}\int_\wp \rho^{(1)}(\bf r \rm) \rho^{(2)}(\bf r \rm-\bf R \rm _{12}) \rho^{(\sigma)}(\bf r \rm) d \bf r \rm
\end{eqnarray}
where $\rho^{(1)}(\bf r\rm)$ is taken from the linearised theory \cite{9,10} and $\rho^{(\sigma)}(\bf r\rm)= \rho_+^{(\sigma)}(\bf r\rm)- \rho_-^{(\sigma)}(\bf r\rm)$ is calculated from eq.(\ref{walldensity}). The correction in eq.(\ref{expansion}) is the lowest order (cubic) term which takes into account the presence of a {\bf charged} surface in the vicinity of the two interacting polyions 1 and 2. There is good reason to believe this correction should be well behaved since for reasonable separations and distances from the wall no two terms will simultaneously be large at any one position in space. Linear theory (which includes terms only up to quadratic order in $F_{id}$), on the other hand, accounts for the dielectric discontinuity at the surface, which reflects itself in the introduction of image charges \cite{9,10}, but is ``blind" to the presence of an electric double layer (i.e.~an inhomogeneous distribution of co and counterions) in the vicinity of the charged surface. An explicit albeit lengthy, expression for $\Delta v_{eff}$ in $\bf k \rm$-space is obtained by taking the Fourier transform of eq.(\ref{expansion}) and invoking the convolution theorem. 

The inverse Fourier transform, to revert to $\bf r \rm$-space, can be reduced to a 1D quadrature which must be carried out numerically and leads to a result of the form;
\begin{eqnarray}
\Delta v_{eff}(h_1,h_2,R_{12})=-k_BT\frac{Z_{eff}^2 \sigma \kappa^4 (n_--n_+) e^{-\kappa h_1}}{ (n_++n_-)^3 }f(h_1,h_2,R_{12})
\end{eqnarray}
where f is a dimensionless function of the arguments $h_1$,$h_2$, the heights of the polyions, and $R_{12}$, the distance between them. In the case where $(h_1=h_2=h)$, f is given by,
\begin{eqnarray}
f(h,R_{12})&=&\int_0^\infty \frac{l J_0(l \kappa R_{12}) dl}{l^2+1}\Bigg(\frac{ 1+g(l)e^{-2(l^2+1)^{1/2}\kappa h}}{2(l^2+1)^{1/2}+ 1} + \frac{1-e^{(1-2(l^2+1)^{1/2})\kappa h}}{2(l^2+1)^{1/2}- 1}\\ \nonumber &+& 2\left(e^{\kappa h}-1 \right) e^{-2(l^2+1)^{1/2}\kappa h}g(l) +\frac{g(l)e^{-2(l^2+1)^{1/2}\kappa h}}{2(l^2+1)^{1/2}+ 1}(1+g(l)e^{\kappa h})\Bigg)
\end{eqnarray}
where,
\begin{eqnarray*}
g(l)=\frac{(l^2+1)^{1/2}-\epsilon_r l}{(l^2+1)^{1/2}+\epsilon_r l}
\end{eqnarray*}
Note that the charge $\sigma$ may be of the same sign as that of the polyions, or of opposite sign. In the latter case the sign of $(n_--n_+)$ will depend on the system. In the canonical ensemble, where the number of ions is well defined by charge neutrality, $(n_--n_+)$ will be negative for a sufficiently high surface charge; while in a semi-grand canonical scheme, where the system is in contact with a reservoir of co and counterions, $(n_--n_+)$ will be determined by the chemical potential of the co  and counterions  in the reservoir. The asympototic form of $\Delta v_{eff}$ for large $R_{ij}$ is,
\begin{eqnarray}
\Delta v_{eff}=-k_BT\frac{Z_{eff}^2 \sigma \kappa^4 (n_--n_+) e^{-\kappa h}}{ (n_++n_-)^3 }\Bigg(\frac{4e^{-\kappa R_{ij}}}{3} &+& \frac{2 e^{-\kappa h}(1-e^{-\kappa h})e^{-\frac{ R_{ij}^2 \kappa}{4h}}}{3h\kappa}\\ \nonumber &+&   \frac{2 \epsilon_r e^{-\kappa h}}{3 \kappa^3 R_{ij}^3}\left(7-4e^{-\kappa h}\right)\Bigg)
\end{eqnarray}
 We have calculated $v_{eff}$ for two polyions situated at an altitude $h$ above the charged plane, as a function of the horizontal distance $d=\mid \bf R \rm _{12} \mid$. Examples of the total effective pair potential (including $\Delta v_{eff}$) are plotted as a function of d in Figures 2 and 3. in Fig. 2 the calculated potentials are shown for several altitudes h, at a fixed surface charge $\sigma =16000 e/ \mu m^{2}$.This surface charge $\sigma$ must once more be considered as an effective charge; the tightly bound counterions of the ``Stern layer'' will not contribute to $\Delta v_{eff}$ in eq.(13). This surface charge is limited to a maximum value of 21000$e/\mu m^{-2}$ by the slit width and the screening length $\lambda_D$ quoted in ref. \cite{4}; the contact-potential corresponding to $\sigma=16000e/\mu m^{-2}$ is 3 in dimensionless units within linear theory    . The physical conditions (polyion size R, charge $Ze$ and concentration $n$; monovalent salt concentration $n_s$ in a closed system) are as close as possible to the experimental conditions of ref. \cite{4} (this reference does not specify the surface charge $\sigma$ of the glass wall which is difficult to measure). As the polyions move closer to the surface, an attractive potential well deepens rapidly and shifts to smaller separations d. Both the depth and position of the potential minimium agree qualitatively with the experimental data \cite{4}, although the calculated attraction appears to be of shorter range. These features vary rapidly with altitude h; at the lower altitudes, when $(h-R)$ becomes comparable to $\lambda_D$ (which is of order of 0.28 $\mu m$ under the present conditions), higher-order terms in the expansion of $F_{id}$ are expected to become important.

In Fig.~3 the same total effective pair potential is plotted versus $d$ for several surface charge densities $\sigma$. The variation with $\sigma$ is seen to be again very rapid; as $\sigma$ increases, the well depth increases and its position shifts to smaller values of $d$. Note that if $\sigma$ is of opposite sign to $Z$; $\Delta v_{eff}$ may still be attractive, since $(n_--n_+)$ may be negative.

In summary we have shown shown that while linearised PB theory invariably predicts a purely repulsive effective interaction between polyions in the bulk or near a surface, perturbative inclusion of the lowest order non-linear term into the ideal contribution to the free energy functional leads to an attractive component of the effective pair potential. The depth of the resulting potential well is of order $k_BT$ under physical conditions  comparable to those of recent experiments [2-4]. Although the attractive component vanishes with $\sigma$, the effect is of partially entropic origin, and is due to a competition of electrostatic and entropic couplings between the ``orbitals''. Since this paper was submitted, a numerical solution of the full non-linear PB theory in a different geometry (two colloidal particles inside a charged cylinder) has been published \cite{15}. This numerical work also predicts an attractive well in the effective polyion-polyion pair potential, in qualitative agreement with the present analytic study.

{\bf Acknowledgements} : the authors are grateful to David Grier for exchange of correspondance concerning his experiments and to Adrian Simper for his advice with some of the mathematics.  DG acknowledges financial support from EPSRC.


%
\begin{figure}
\label{figure1}
\centerline{\psfig{figure=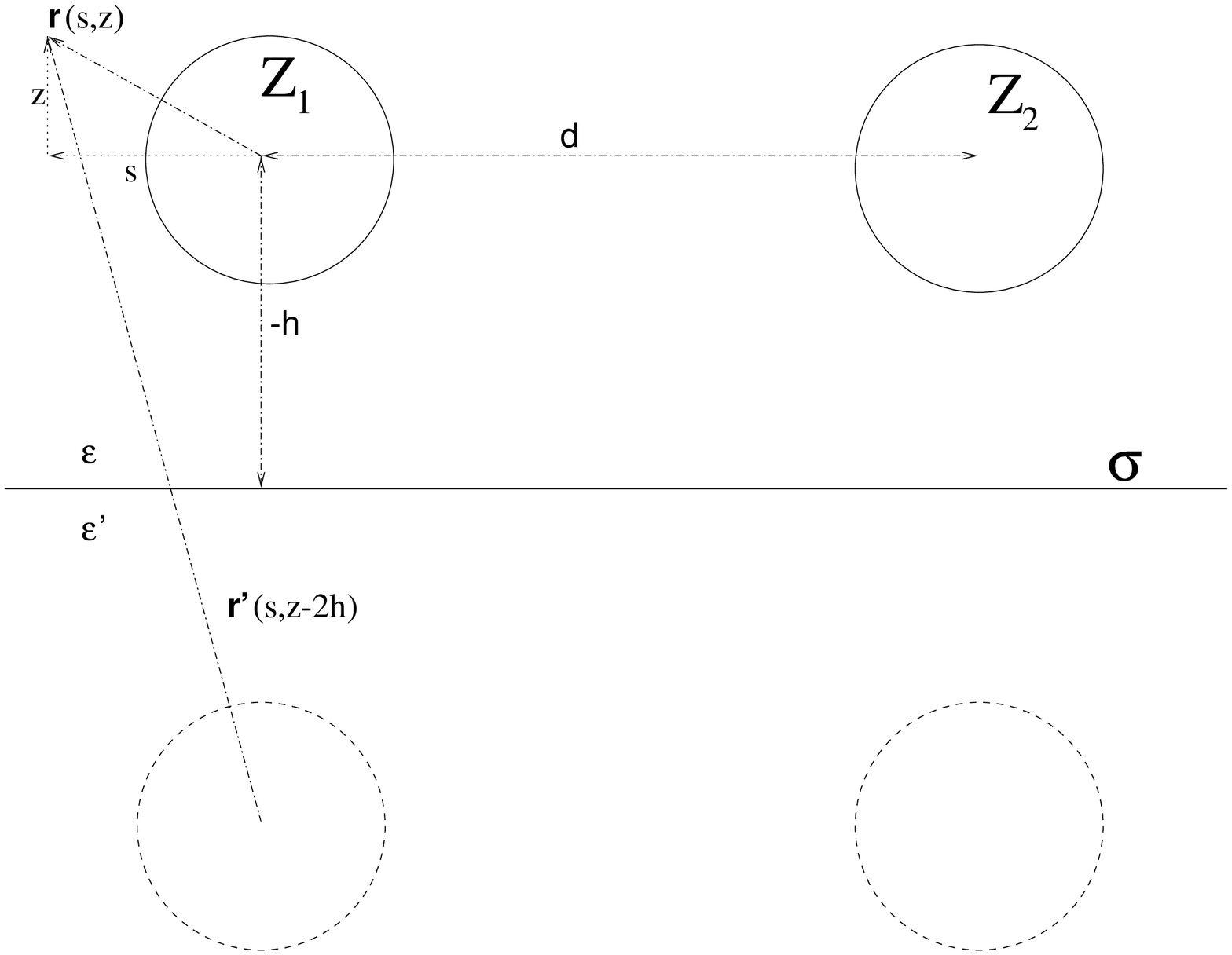,width=\textwidth}}

\caption{Geometry of particles near a planar surface. The suspension, of dielectric constant  $\epsilon$, is confined to $z>-h$, while a material of dielectric constant $\epsilon'$ occupies the half-space $z<-h$ and carries a surface charge $\sigma$. The dotted circles represent the image charges.}
\end{figure}
\begin{figure}
\label{figure2}
\centerline{\psfig{figure=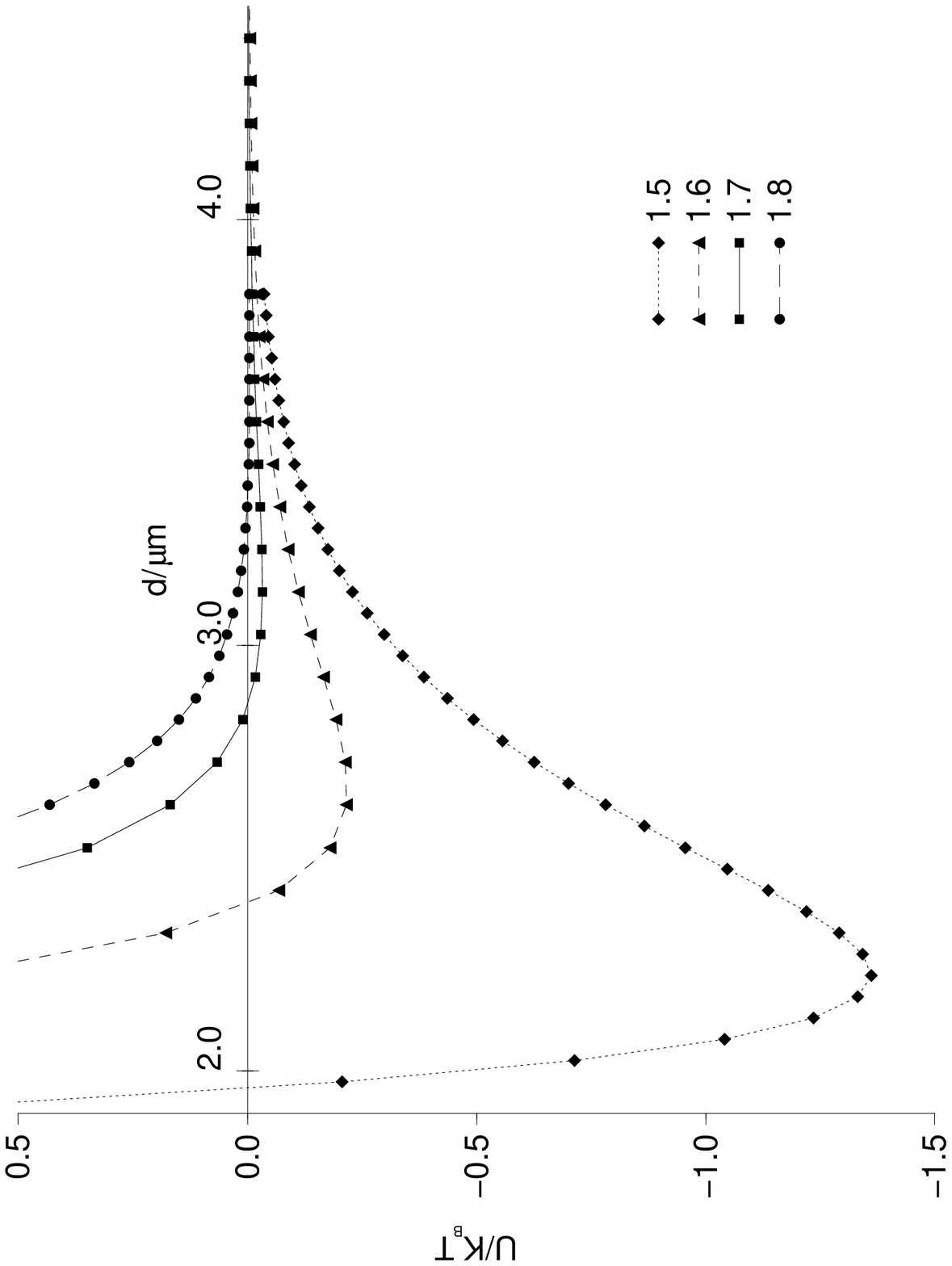,width=\textwidth}}
\caption{  Dimensionless total (Linear + Perturbative solution) effective pair potential, $v_{eff}(h;d)/k_BT$ versus distance d between polyions situated at the heights shown in the legend, measured in $\mu$m. The polyions have a bare valence Z=5000, and a diameter $D=0.65\mu m$. $T=298K$; $\epsilon=78$ and $\epsilon'=6.3$; $n_s=3\times10^{-7}M$ and the wall carries a charge of 16000 e/$\mu m^2$}
\end{figure}
\begin{figure}
\label{figure3}
\centerline{\psfig{figure=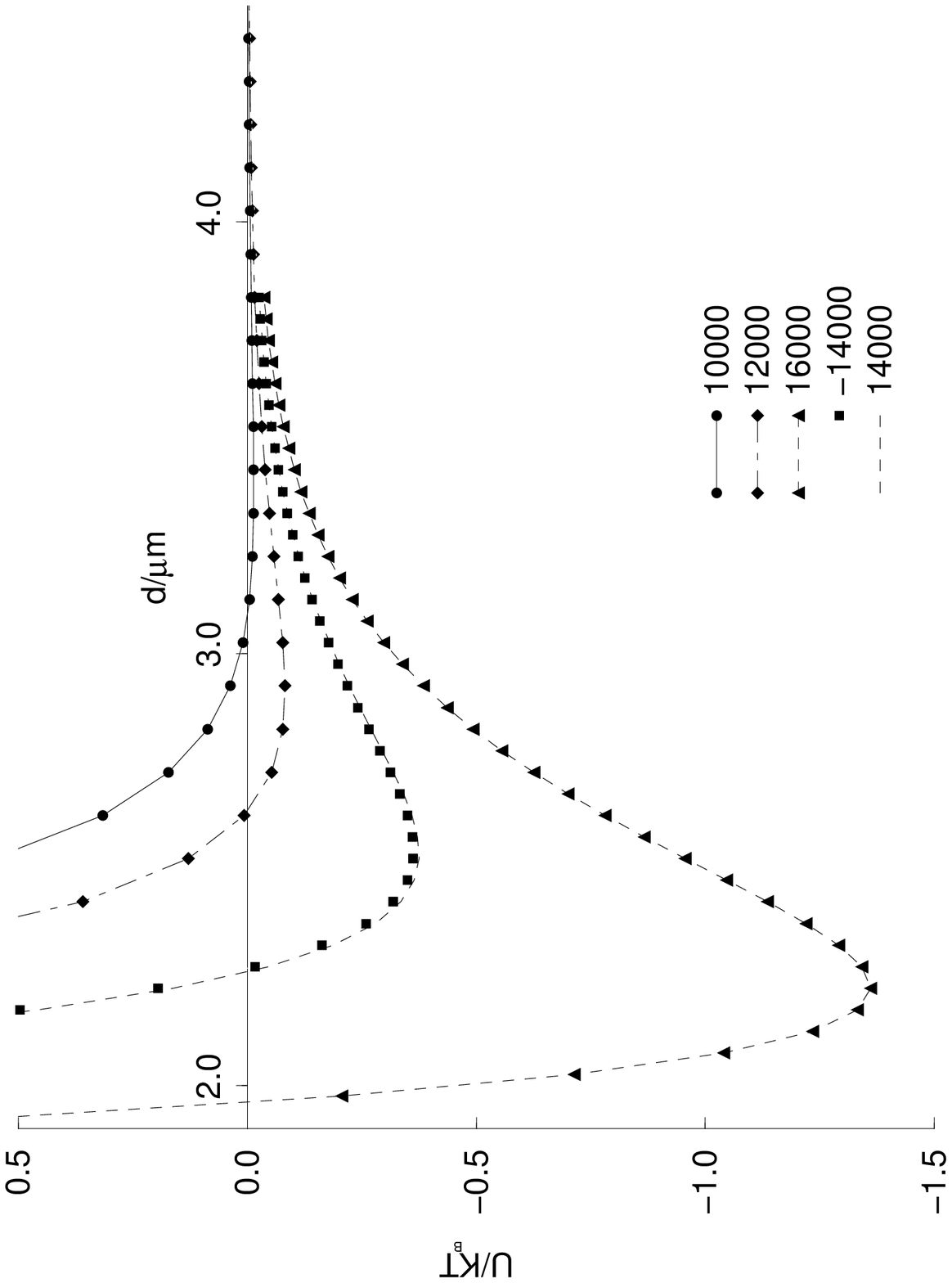,width=\textwidth}}
\caption{Same as in figure 2 but this time the height of the particles is fixed at 1.5 $\mu m$ and the surface charge is varied as shown and is measured in $e/\mu m^2$; the Debye length is fixed at $\lambda_D=0.28\mu m$, so that the salt concentration $n_s$  varies with $\sigma$}
\end{figure}
%

\end{document}